# Solar-Like Oscillations in a Massive Star

Kévin Belkacem,[1,2]* Réza Samadi,[1] Marie-Jo Goupil,[1] Laure Lefèvre,[1] Fréderic Baudin,[3] Sébastien Deheuvels,[1] Marc-Antoine Dupret,[1,2] Thierry Appourchaux,[3] Richard Scuflaire,[2] Michel Auvergne,[1] Claude Catala,[1] Eric Michel,[1] Andrea Miglio,[2] Josefina Montalban,[2] Anne Thoul,[2] Suzanne Talon,[4] Annie Baglin,[1] Arlette Noels[2]

Seismology of stars provides insight into the physical mechanisms taking place in their interior, with modes of oscillation probing different layers. Low-amplitude acoustic oscillations excited by turbulent convection were detected four decades ago in the Sun and more recently in low-mass main-sequence stars. Using data gathered by the Convection Rotation and Planetary Transits mission, we report here on the detection of solar-like oscillations in a massive star, V1449 Aql, which is a known large-amplitude (β Cephei) pulsator.

Stars burn hydrogen into helium through nuclear fusion during most of their life. Once the central hydrogen gets exhausted, the helium core starts contracting, and hydrogen-shell burning takes over as the main energy source. The subsequent evolution depends mostly on a star's mass at birth but also on the physical mechanisms occurring during the hydrogen-burning phase. For instance, transport of chemical elements determines the helium core size, which is crucial for the evolution of stars. Transport processes such as turbulence and those induced by rotation are not fully understood and are still poorly modeled, but stellar seismology can provide important constraints provided that the modes that probe the relevant regions are excited, detected, and identified. This will be the case for stars oscillating over a large range of oscillation modes probing different layers of the star.

Here, we report on the detection of solar-like oscillations (high-frequency acoustic modes that are damped but excited by turbulent convection and probe superficial convective regions) in a 10–solar mass star, V1449 Aql, already known to be a β Cephei (it oscillates on unstable low-frequency modes of high amplitude, also referred to as opacity-driven modes, which probe the deepest regions of stars) (*1*).

The largest-amplitude mode in the Fourier spectrum of V1449 Aql has been detected from the ground (*1*, *2*) at a frequency of 63.5 μHz. Those pulsations are excited by a thermal instability known as the κ-mechanism (*3*), which in the present case is related to the existence of an iron-opacity bump located in the upper layers of massive stars. The iron-opacity bump in such a massive star induces the existence of a convective zone in the upper layers (*4*), which could be responsible for the excitation of the detected modes.

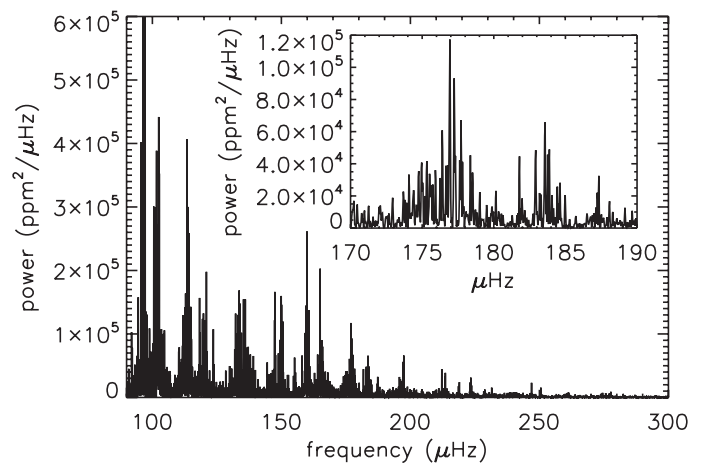

**Fig. 1.** Fourier spectrum of prewhitened light curve obtained from the quasi-uninterrupted 150 days of observations, with a duty cycle of 90%, of the star V1449 Aql by CoRoT, showing structures that are reproduced over the 100- to 250-μHz interval. (**Inset**) Enlarged part of the spectrum showing a typical solar-like structure. Below 100 μHz, we enter the bulk regime of unstable modes (fig. S1), and the possible existence of many such modes in this frequency domain then makes the deciphering of unstable versus stable modes quite delicate. Hence, to remain conservative we restrict the discussion to frequencies above 100 μHz.

[1]Laboratoire d'Études Spatiales et d'Instrumentation en Astrophysique, CNRS (UMR 8109), Observatoire de Paris, Place J. Janssen, F- 92195 Meudon, France; associated with Université Pierre et Marie Curie and Université Denis Diderot. [2]Institut d'Astrophysique et de Géophysique de l'Université de Liège, Allée du 6 Août 17–B 4000 Liège, Belgium. [3]Institut d'Astrophysique Spatiale, Université Paris-Sud 11 and CNRS (UMR 8617), Batiment 121, F-91405 Orsay, France. [4]Réseau Québécois de Calcul de Haute Performance, Université de Montréal, Casier Postal 6128, Succursale Centre-ville, Montréal, Québec H3C 3J8, Canada.

*To whom correspondence should be addressed. E-mail: kevin.belkacem@ulg.ac.be

Our results are based on the quasi-uninterrupted (meaning, a duty cycle of 90%) light curve obtained over 150 days with the Convection Rotation and Planetary Transits (CoRoT) (*5–7*) Centre National d'Etudes Spatiales (CNES) space mission (Fig. 1). The dominant opacity-driven mode is located at 63.5 µHz with an amplitude of $3.9 \times 10^4$ parts per million (ppm). We looked for stochastically excited modes with frequencies above 100 µHz; below this limit, the existence of several opacity-driven modes makes the analysis more difficult. No signal is found above 250 µHz. In the frequency range 100 to 250 µHz, there are broad structures with a width of several µHz, with low amplitudes of hundreds of parts per million and well above the noise level, which is around 1 ppm.

These structures are not the result of instrumental effects [supporting online material (SOM) text]. To verify that they are not related to the existence of the opacity-driven modes, we carried out prewhitening (SOM text), which suppresses the influence of the dominant peaks and their harmonics in the relevant frequency domain as well as related aliases because of the observational interruptions. We validated our prewhitening method through numerical simulations (SOM text).

The modes associated with the broad structures have a finite lifetime, in contrast with opacity-driven modes, which are coherent oscillations and therefore appear as sinus cardinal functions in the Fourier spectrum. The amplitudes of these oscillations vary stochastically in time, again in contrast with the stationary property of the dominant opacity-driven mode and its harmonics. Their power is intermittent in time and dispersed in terms of frequency (Fig. 2). Such behavior, typical of solar-like modes (*8*), confirms the stochastic nature of the structures. In contrast, the temporal behavior of the second harmonic of the fundamental opacity-driven mode, which lies in the same frequency interval, is centered on a single frequency (Fig. 2, bottom). The widths of the detected high-frequency modes show that they are damped, which is a signature of modes excited by turbulent convection.

We looked for regularly spaced patterns in the Fourier spectrum, which are a characteristic signature of those modes. An autocorrelation of the Fourier spectrum shows periodicities centered around 5, 14, and 27 µHz (Fig. 3), indicating the existence of periodicities in the power spectrum.

Theoretical calculations show that these properties, interpreted as damped acoustic modes excited by turbulent convection, are compatible with solar-like oscillations of a massive main sequence star. We carried out numerical simulations using a 10–solar mass stellar model that is appropriate for V1449 Aql in that it corresponds to the observational constraints obtained from ground-based observations (*9*). A comparison between the theoretical and observational autocorrelations shows that the observed frequency spectrum is compatible with the presence of modes of angular degrees l = 0, 1, and 2, characterized by a large frequency separation around 27 µHz, with 1-µHz widths and a rotational splitting of 2.5 µHz related to a rotation with an axis inclined by 90° with respect to the line of sight (SOM text).

Mode amplitudes obtained from theoretical computations of the line width (*10*) and the energy supplied in the mode by turbulent convection (*11*) reach several tens of parts per million, which is well above the CoRoT detection threshold and in agreement with observations. Our calculations show that excitation by the turbulent convective motions associated with the iron-opacity bump in the upper layers of the star is efficient. This driving is operative when the convective time scale of energy-bearing eddies is close to the modal period, which explains why modes in the frequency range of 100 to 250 µHz are observed.

In summary, we showed that the broad structures at high frequencies detected in the CoRoT Fourier spectrum of the star V1449 Aql are not the result of instrumental effects, are independent of the opacity-driven modes, and present regularly spaced patterns that are characteristic of high-frequency acoustic modes. These structures have the theoretically expected properties of solar-like oscillations: modes excited by turbulent convection.

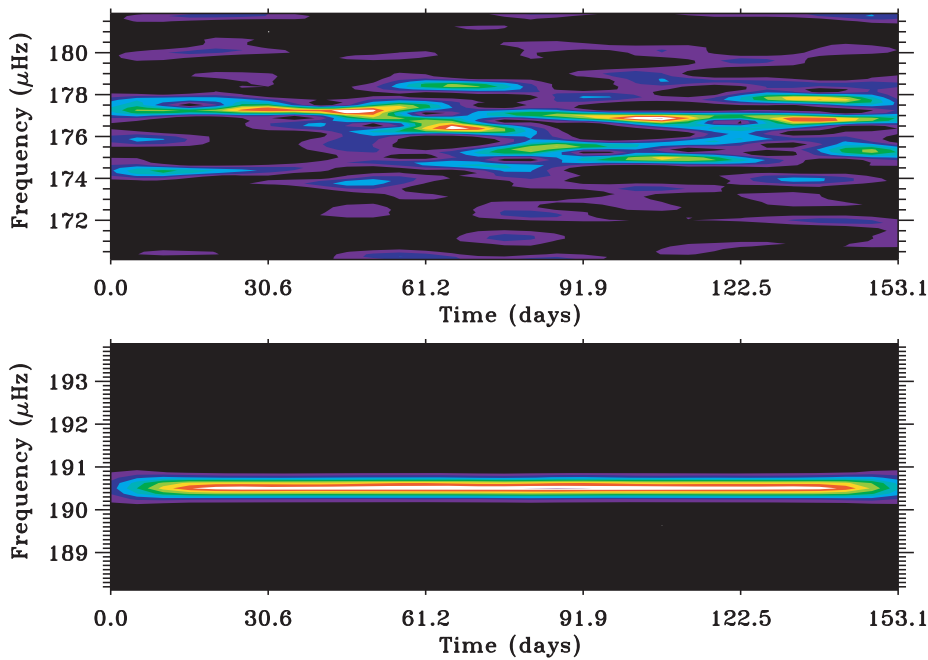

**Fig. 2.** Time-frequency diagram, using a Morlet wavelet with a 20-day width (*8*). (**Top**) Solar-like mode in the prewhitened light curve shown in the inset of Fig. 1, which exhibits a time-dependent behavior and a spreading over several µHz. (**Bottom**) For comparison, the second harmonic of the dominant peak, associated with the opacity-driven mode in the unprewhitened light curve.

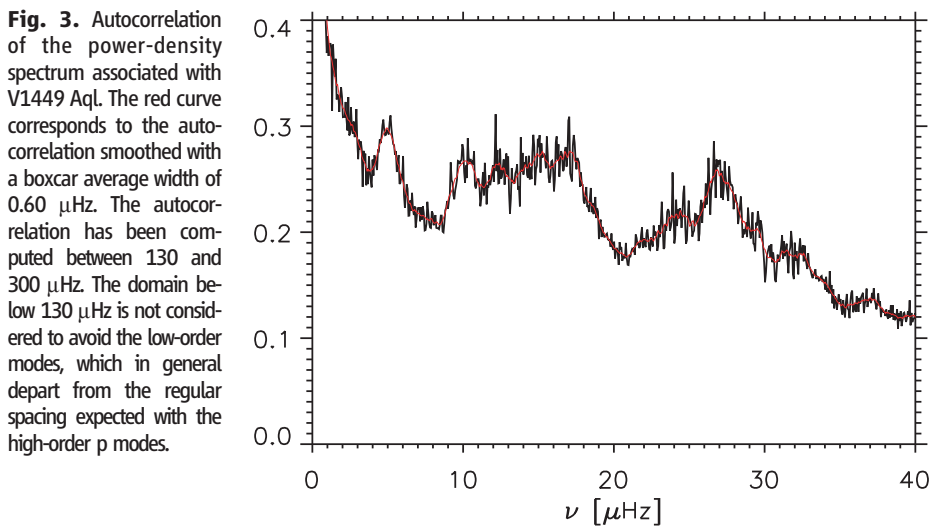

**Fig. 3.** Autocorrelation of the power-density spectrum associated with V1449 Aql. The red curve corresponds to the autocorrelation smoothed with a boxcar average width of 0.60 µHz. The autocorrelation has been computed between 130 and 300 µHz. The domain below 130 µHz is not considered to avoid the low-order modes, which in general depart from the regular spacing expected with the high-order p modes.

# Supporting online material

# Solar-like oscillations in a massive star

Kévin Belkacem[1,2], Réza Samadi[1], Marie-Jo Goupil[1], Laure Lefèvre[1], Fréderic Baudin[3], Sébastien Deheuvels[1], Marc-Antoine Dupret[1,2], Thierry Appourchaux[3], Richard Scuflaire[2], Michel Auvergne[1], Claude Catala[1], Eric Michel[1], Andrea Miglio[2], Josefina Montalban[2], Anne Thoul[2], Suzanne Talon[4], Annie Baglin[1], Arlette Noels[2]

[1] LESIA, CNRS (UMR 8109), Observatoire de Paris, pl. J. Janssen, F- 92195 Meudon, France, associé Université Pierre et Marie Curie, Université Denis Diderot
[2] Institut d'Astrophysique et de Géophysique de l'Université de Liège, Allée du 6 Août 17 – B 4000 Liège, Belgium
[3] Institut d'Astrophysique Spatiale (IAS), Université Paris-Sud 11 and CNRS (UMR 8617), Batiment 121, F-91405 Orsay, France
[4] Réseau québécois de calcul de haute performance (RQCHP), Université de Montréal, C.P. 6128, succursale centre-ville, Montréal, Québec, H3C 3J8

## I) Data analysis

**a) Instrumental effects**

As a first test to verify that the structures seen in the frequency domain [100; 250] µHz of the Fourier spectrum of V1449 Aql (HD180642) are not due to some instrumental perturbations, we have considered the fainter (apparent visual magnitude mv=9.14 for HD181072, and mv=8.27 for V1449 Aql (*S1*)) CoRoT target HD181072, which is in the same observation run and same CCD.
HD181072 does not exhibit structures comparable to those observed in V1449 Aql in the abovementioned frequency region. Its frequency spectrum only shows a residual of the orbital frequency with an amplitude around 9.8 ppm and a flat/white noise with an amplitude of 0.6 ppm in 150 days, far below the amplitudes of the structures we are interested in. Furthermore, neither the star background nor the satellite jitter show the structures observed in the stellar V1449 Aql.

**b) Prewhitening procedure**

To remove the influence of the dominant peaks and associated perturbation due to observational interruptions, we performed a prewhitening of the light curve (hereafter LC) obtained by CoRoT (Fig *S1*). The highest peaks in the time-series are fitted and then removed from the time series. Within each frequency range of 2 µHz, peaks are selected whenever their amplitudes are at least five times the local mean of the power spectrum. A fit to the time series is performed, based on a Levenberg-Marquardt method with a least-square minimization (*S2, S3, S4*). A total of 91 peaks were withdrawn. We find that the large structures in the power spectrum, which we identify as solar-like modes, are unaltered by the prewhitening procedure; their existence is therefore independent of the low frequency opacity-driven modes.

**c) Validation using numerical simulations**

In order to validate our prewhitening procedure as well as to theoretically investigate the influence of the combined effects of the unstable dominant modes and the observational interruptions, we built two simulated LC using our CoRoT light-curve simulator (*S5*). The first simulation is a temporal signal built by including only the low-frequency unstable modes that have been detected in the LC of V1449 Aql down to the amplitude of 460 ppm. In the second simulation, we also included a set of solar-like oscillations. The frequencies and line widths of these solar-like oscillations are computed using the non-adiabatic numerical code MAD (*S6*) and a stellar model- built with the evolutionary numerical code CLES (*S7*) - that matches the effective temperature and the luminosity of the star (see section II for details). Amplitudes of the included solar-like modes are set to a level representative of the high-frequency stable modes detected in V1449 Aql, that is 300 ppm. Each simulated LC is sampled with the same duty-cycle than that associated with the observations of V1449 Aql and noise is added to the resulting time series. Finally, we filled the gaps in the same way as done by the CoRoT data treatment chain (*S8*).

The prewhitened Fourier spectrum associated with the first simulated LC does not show any structures in the frequency domain we are interested in. This demonstrates that the combined effects of the unstable dominant modes, the window and the filling procedure do not produce spurious structures that could erroneously be interpreted as solar-like oscillations.

The second LC, which includes solar-like modes, has been prewhitened to remove the unstable modes in the same manner than done for V1449 Aql. The Fourier spectrum associated with the second simulated LC is presented in Fig. S2. The simulated solar-like oscillations are recovered in the Fourier spectrum associated with the second simulated LC. This demonstrates that the clean procedure leaves existing solar-like oscillations unaltered.

## II) Modelling of solar-like modes

### a) Auto-correlation of HD180642

Solar like oscillations are acoustic oscillations whose frequencies are expected to follow more or less closely the asymptotic regime (*S9*). In order to search for such a signature of regular p-mode spacing, we computed the autocorrelation of the Fourier spectrum in the frequency domain 130 µHz – 300 µHz. The domain below 130 µHz is excluded so as to avoid the low-frequency (low-radial order) modes, which in general significantly depart from the regular spacing expected for the high frequency p modes. The result is shown in Fig. S3, which clearly reveals three patterns, one centred on $\nu_0 = 5$ µHz, the second around $\nu_1 = 13.5$ µHz and the last one around $\nu_2 = 27$ µHz. The second pattern is centered at a frequency equal to half $\nu_2$, that is $\nu_1 = \nu_2/2$. The autocorrelation function then shows that a regular frequency spacing exists in the Fourier spectrum of the light-curve of V1449 Aql. This is the expected signature of asymptotic p modes with an associated large separation of either $\nu_2 = 27$ µHz or half its value, $\nu_1$. We also see that the peak at $\nu_2$ in the autocorrelation function has several components that are distant from $\nu_2$ by a multiple of 2.5 µHz. The same feature, although less pronounced, is also observed in the pattern centred on $\nu_1$. The presence of the peak at $\nu_0$=5 µHz can be interpreted as the signature of mode splitting induced by the rotation of the star.

**b) Simulated auto-correlation**

An important issue is to verify that the properties of high frequency p modes, which are derived from the observed spectrum, are -at least roughly- consistent with theoretical expectations for solar-like p modes in a massive star. We therefore carried out a modelling of V1449 Aql, which is not fully optimized but sufficient for the present purpose. Future works will be dedicated to an optimized modelling of V1449 Aql.

To simulate an autocorrelation that can be compared with the one derived from the observed V1449 Aql LC, we computed a model that approximately matches the effective temperature of the star (*S1*) ($T_{eff}$=24500 ± 1000 K, log g = 3.45 ± 0.15). A ten-solar-mass model is obtained, with an effective temperature of 24075 K and a gravity of log g=3.88, using the CLES evolutionary code (*S7*). The associated p-mode frequencies present a large separation of 27 µHz, as suggested by the observed autocorrelation (Fig. S3). In building the simulated light curve, we included modes with angular degree l=0, 1, 2 and 3. We assumed the same theoretical value, $\sigma$=1 µHz, obtained with the code MAD (*S6*) for all line-widths of the modes. The radial modes were all given amplitude of 300 ppm, which is representative of the structures observed for V1449 Aql. The amplitude of the non-radial modes were set up to a 300 ppm value modulated by mode visibility factors. We further assumed a constant rotational splitting $\sigma$ and defined *i* as the inclination angle between the rotation axis and the line of sight. From our set of theoretical frequencies, we simulate a LC using the CoRoT light-curve simulator[15] and computed the associated power density spectrum and the autocorrelation function.

We investigated different values of $\sigma$ and *i*. An agreement with the observations is found for a rotational splitting of $\sigma$= 2.5 µHz and an inclination angle of *i*=90°. The result is shown in Fig S4. The pattern at $\nu_0$ is reproduced. Our theoretical result shows that including p-modes split by a rotation of 2.5 µHz with an inclination angle of i=90° gives an autocorrelation that is compatible with the observed one. In order for the reader to estimate the quality of the agreement, we also show two additional autocorrelation functions in Fig. S5. In the first case we assume i=45° and a splitting of 2.5 µHz (top panel), while in the other case we assume i=90° with no splitting, i.e. keeping only the central component of the rotationally split multiplets (bottom panel). These autocorrelation functions do not reproduce the observed structures seen in Fig. S3.

These results provide a theoretical support to our assumption that the structures observed on the autocorrelation function of V1449 Aql are compatible with the presence of a regular p-mode spectrum. We also find that the main features of the observed autocorrelation function can be reproduced by assuming an internal rotation of 2.5 µHz and a rotation axis that has an inclination of i~ 90° with respect to the observed direction. However, a definite determination of the values of angle i and the rotation frequency is dedicated to a forthcoming work together with an optimized modeling.

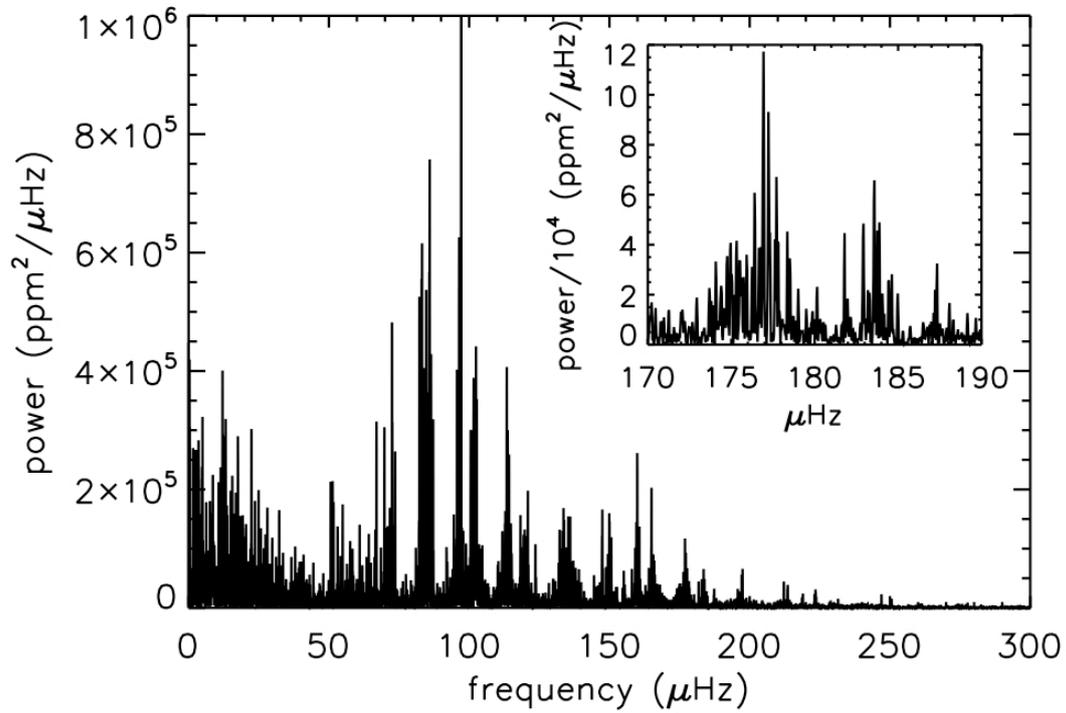

**Fig S1:** Fourier spectrum of prewhitened light curve obtained from the quasi-uninterrupted 150 days of observations, with a duty-cycle of 90%, of the star V1449 Aql by CoRoT, showing structures that are reproduced. Below 100 µHz, we enter the bulk regime of unstable modes and the possible existence of many such modes in this frequency domain then makes the deciphering of unstable versus stable modes quite delicate. Below 50 µHz, one notes an increase of the power toward low frequencies that is not yet identified as due to coloured noise, and/or g modes, and/or rotation. (Inset) Enlarged part of the spectrum.

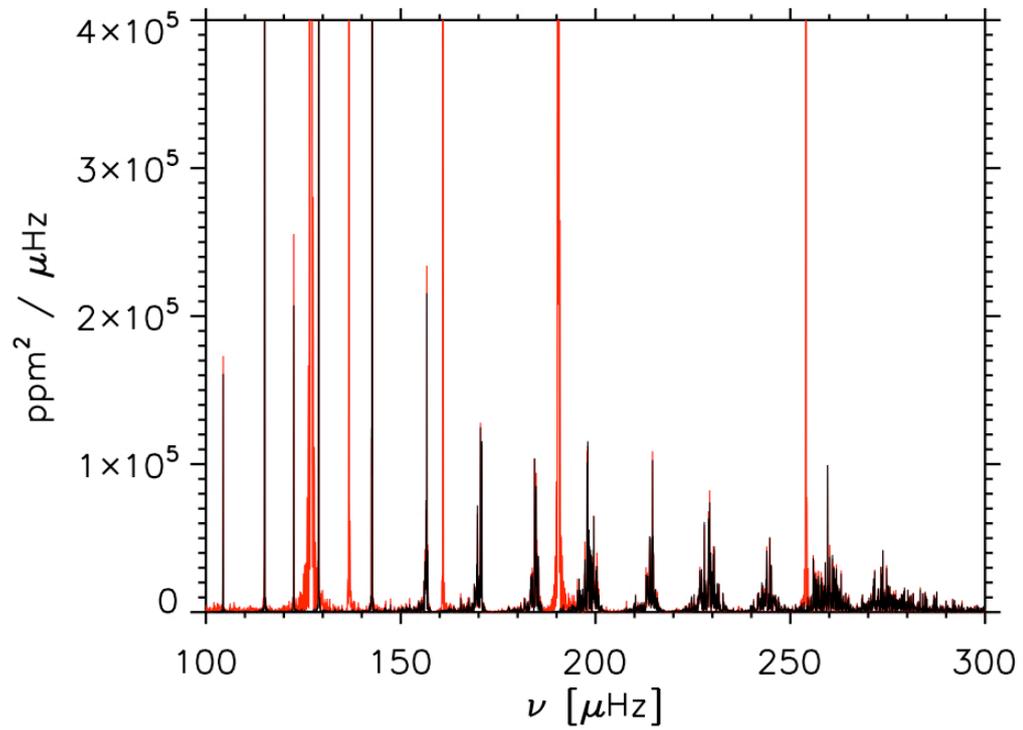

**Fig S2:** Power spectral density of the *simulated* LC that includes both unstable modes and solar-like oscillations. In black: prewhited LC. In red: original LC

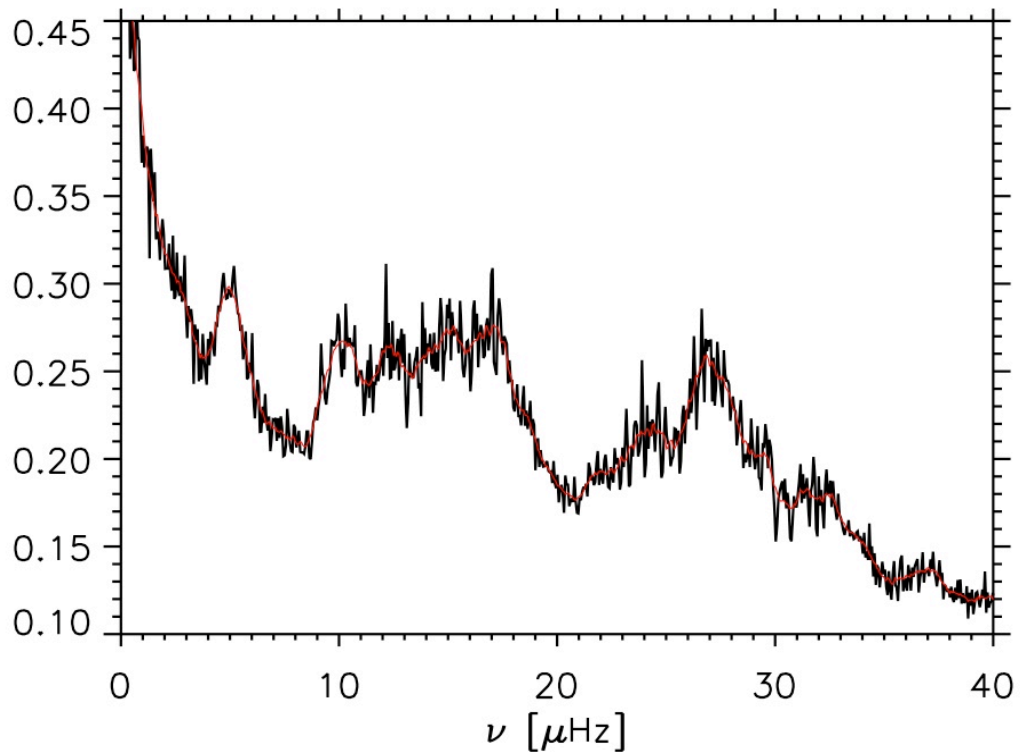

**Fig S3:** Autocorrelation of the power density spectrum associated with the observed lightcurve of V1449 Aql. The autocorrelation has been computed between 130 µHz and 300 µHz.

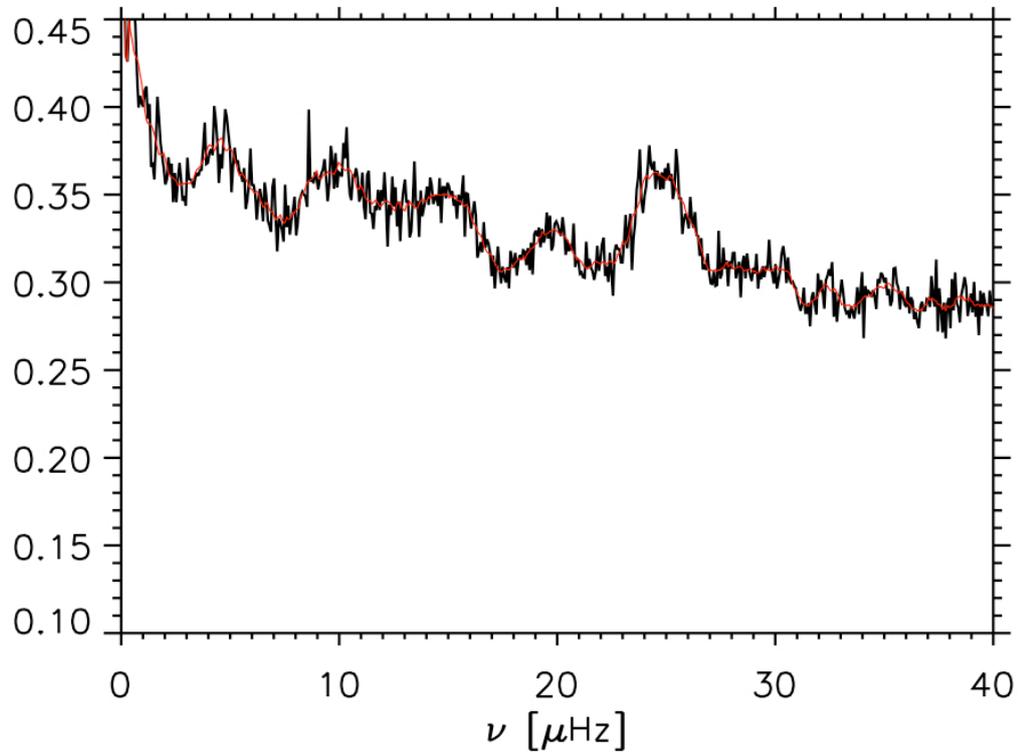

**Fig S4:** Autocorrelation function of the *simulated* power density spectrum associated with the stellar model that matches the effective temperature of V1449 Aql and that has a large separation of 27 µHz. It includes the modes l=0,1 and 2, with i=90° and a splitting of 2.5 µHz.

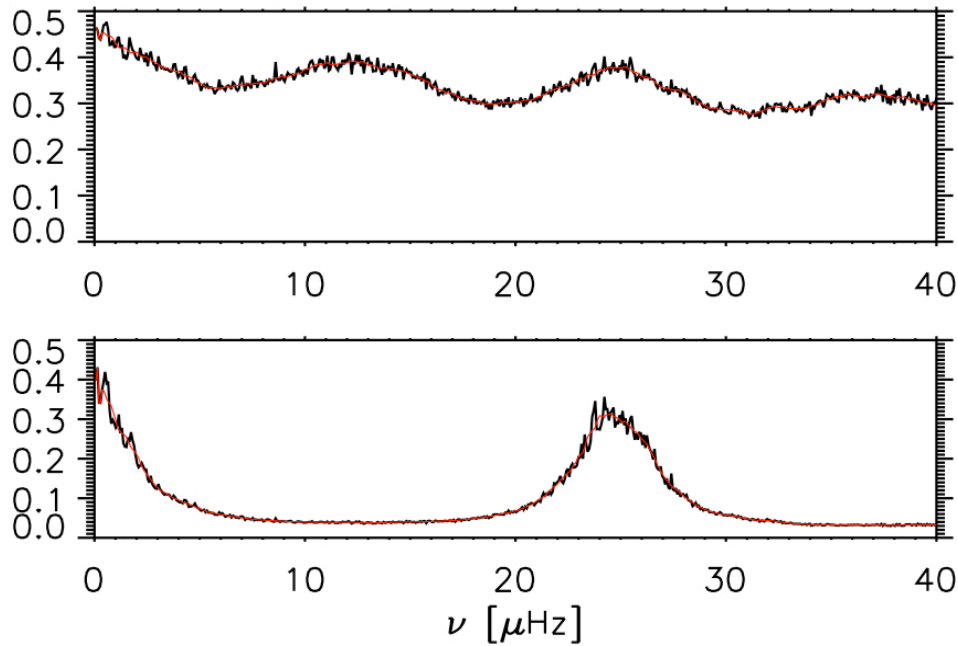

**Fig S5:** Autocorrelation function of simulated power density spectrum associated with the stellar model that matches the effective temperature of V1449 Aql and that has a large separation of 27 μHz. At the top panel, the simulation includes the modes l=0,1 and 2, with i=45° and a splitting of 2.5 μHz. At the bottom panel, the simulation includes the modes l=0,1 and 2, with i=90° and no splitting.